\documentclass[twocolumn,showpacs,amsmath,amssymb,aps]{revtex4-1}
\usepackage{graphicx}
\usepackage{multirow}
\usepackage{bm}
\usepackage[breaklinks=true,colorlinks=true,linkcolor=blue,urlcolor=blue,citecolor=blue]{hyperref}
\usepackage{booktabs}
 \usepackage{color}
 \usepackage[normalem]{ulem} 

\begin{document}

\title{Spin-Valley-Mismatched Altermagnet for Giant Tunneling Magnetoresistance}

 \author{Kun \surname{Yan}$^{1}$}
 \author{Yizhi \surname{Hu}$^{1}$}
 \author{Wei-Hua \surname{Xiao}$^{1}$}
 \author{Xiaolong \surname{Zou}$^{2}$}
 \email{Email:xlzou@sz.tsinghua.edu.cn}
 \author{Xiaobin \surname{Chen}$^{1,3}$}
 \email{Email:chenxiaobin@hit.edu.cn} \author{Wenhui \surname{Duan}$^{4,5,6}$}

\affiliation{
$^1$School of Science, State Key Laboratory on Tunable Laser Technology and Ministry of Industry and Information Technology Key Lab of Micro-Nano Optoelectronic Information System, Harbin Institute of Technology, Shenzhen, Shenzhen 518055, China\\
$^2$Shenzhen Geim Graphene Center \& Shenzhen Key Laboratory of Advanced Layered Materials for Value-added Applications, Institute of Materials Research, Tsinghua Shenzhen International Graduate School, Tsinghua University, Shenzhen 518055, China\\
$^3$Collaborative Innovation Center of Extreme Optics, Shanxi University, Taiyuan 030006, China\\
$^4$State Key Laboratory of Low Dimensional Quantum Physics, Department of Physics, Tsinghua University, Beijing 100084, China\\
$^5$Frontier Science Center for Quantum Information, Beijing 100084, China\\
$^6$Institute for Advanced Study, Tsinghua University, Beijing 100084, China
}

\date{\today}

\begin{abstract}
\textbf{Abstract} Altermagnet-based heterojunctions have demonstrated magnetoresistive effects in experiments, however, a predictive theoretical model for non-ferromagnetic structures has remained elusive. In this work, we develop a tunneling-based spin-transport theory that explicitly incorporates the transverse-wavevector ($\bf{k}_\|$)-dependent spin polarization of an altermagnet's transport channels, enabling the prediction of giant tunneling magnetoresistance (TMR). Based on the theory, we predict that the altermagnet KV$_2$Se$_2$O can reach the extreme limit of magnetoresistance. By performing first-principles transport calculations, we verify that magnetic tunnel junctions using the metallic KV$_2$Se$_2$O as the electrodes and few-layer MgO as the spacer exhibit zero-bias magnetoresistance larger than $7.57\times10^7$\%, which is robust against the bias and thickness of the spacer. Our research provides a quantitative design principle for next-generation spin-electronic devices and establishes KV$_2$Se$_2$O/MgO/KV$_2$Se$_2$O as a leading candidate material system for room-temperature ultra-high-density non-volatile memory.

\end{abstract}
\maketitle

\section*{Introduction}
Magnetic devices with high magnetoresistance (MR) can help address a range of major technological challenges, such as improving energy efficiency and enabling device miniaturization. A magnetic tunnel junction (MTJ) generally comprises two metallic ferromagnetic materials (FM), which are separated by a nonmagnetic insulating material serving as a tunnel barrier~\cite{julliere1975tunneling,yuasa2004giant,moodera1995large}. The resistance of an MTJ changes with the alignment of the magnetic moments of the two ferromagnetic materials, which can be parallel (P) or antiparallel (AP). The optimistic/pessimistic definition of MR is
\begin{equation}
    \mathrm{MR}_\textrm{o/p} = \frac{R^{\textrm{AP}}-R^{\textrm{P}}}{R^{\textrm{P/AP}}}\times 100\%,
\end{equation}
where $R^{\textrm{AP/P}}$ is the resistance under AP/P configuration. This magnetoresistive phenomenon is commonly understood to arise from the spin-polarized tunneling current modulated by the relative magnetization orientations of the ferromagnetic electrodes and typically quantified using Julliere's formula: $\mathrm{MR}_\textrm{o/p}=2 \eta_1\eta_2/(1\mp \eta_1 \eta_2)$, where $\eta_{1}$ and $\eta_{2}$ represent the spin polarization of the charge carriers on two electrodes~\cite{wang2015spin,zhao2017magnetic}. According to this formula, enhancing the spin polarization of the electrodes results in higher MR. Notably, this formula does not apply to antiferromagnetic electrodes because the spin polarization of the charge carriers in an antiferromagnetic material is zero, and therefore the MR should be zero~\cite{jiang2023prediction}.

\begin{figure}[htp]
 \centerline{\includegraphics[width=0.45\textwidth]{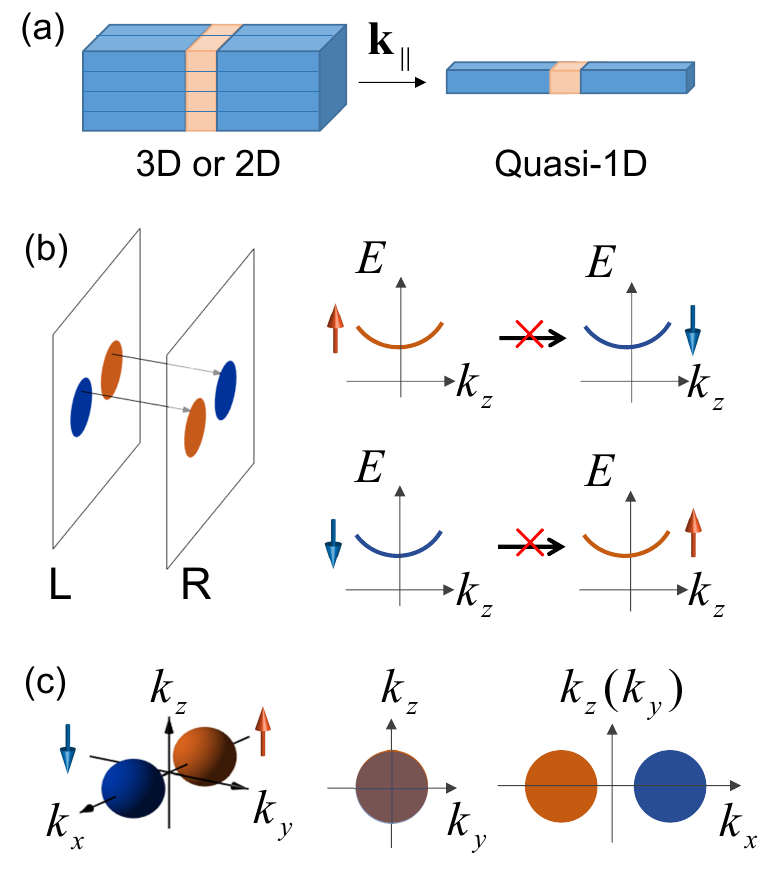}}
 \caption{\textbf{Schematic plots of spin-valley-mismatched transport theory.} (a) A 3D transport junction can be deemed as a quasi-1D model for each transverse wave vector $\bf{k_\parallel}$. (b) If the spin-resolved transport channels of the left (L) and right (R) leads do not overlap, the charge transport under the antiparallel configuration is blocked. (c) For the isosurface of band energies, the distribution of transport channels is the projection of the isosurfaces on the transverse plane. Shaded areas in the middle panel and right panel indicate two and one transport channels, respectively. }\label{theory}
 \end{figure}
Recent progress suggests that altermagnets, which are special antiferromagnets featuring spin-split band structures, encompassing non-collinear~\cite{naka2019spin,vzelezny2017spin,gurung2021transport} and specific collinear configurations~\cite{vsmejkal2022beyond,vsmejkal2022emerging}, can sustain longitudinal spin-polarized currents and instigate magnetoresistive response~\cite{bai2022observation}.
Theoretically, MTJs based on the altermagnet RuO$_2$ have been reported to exhibit a magnetoresistance as high as 500\%  in RuO$_2$/TiO$_2$/RuO$_2$(001)~\cite{shao2021spin}, 200\%  in RuO$_2$/TiO$_2$/RuO$_2$(110)~\cite{jiang2023prediction}, and 6100\% in RuO$_2$(110)/TiO$_2$/CrO$_2$~\cite{chi2024crystal}.
There are also several experimental reports on the MR of RuO$_2$-based MTJs. A magnetoresistance of 60\% at room temperature was observed in RuO$_2$/MgO/RuO$_2$~\cite{xu2023spin}, and 5\% was observed at 10 K in RuO$_2$/TiO$_2$/CoFeB MTJ~\cite{Noh2025_RuO2TMR}. Despite of the rapid developments of antiferromagnet-based resistive devices, which have the potential advantage of electrical switching of altermagnetism~\cite{Wadley_Science2016,LuHaizhou_PRL2025,Taoll_PRL2024,fu2025}, an intuitive and predictive model for estimating MR is still lacking, and achieving extreme-limit MR using antiferromagnets remains challenging~\cite{SmejkalLibor_PRX2022}.

In this work, we propose a general formula for the estimation of MR in MTJs, which can be composed of ferromagnets or antiferromagnets. Based on the formula, we predict that materials have an effective spin polarization of 100\% would exhibit extremely large MR. By combining with first-principle transport calculations, we further demonstrate that the metallic altermagnet KV$_2$Se$_2$O is a spin-valley-mismatched altermagnetic material, which has an effective spin polarization of 99.93\%. Utilizing KV$_2$Se$_2$O as electrodes and $l$-layer MgO as the spacer, we show that giant zero-bias MR of more than $7.57\times10^{7}$\% (optimistic) and 99.999\% (pessimistic) emerges. Our findings shed light on the direction for searching for high-MR devices in generalized MTJs and reveal the vast potential of altermagnet KV$_2$Se$_2$O in MR devices.

\begin{figure*}[htp]
 \centerline{\includegraphics[width=0.85\textwidth]{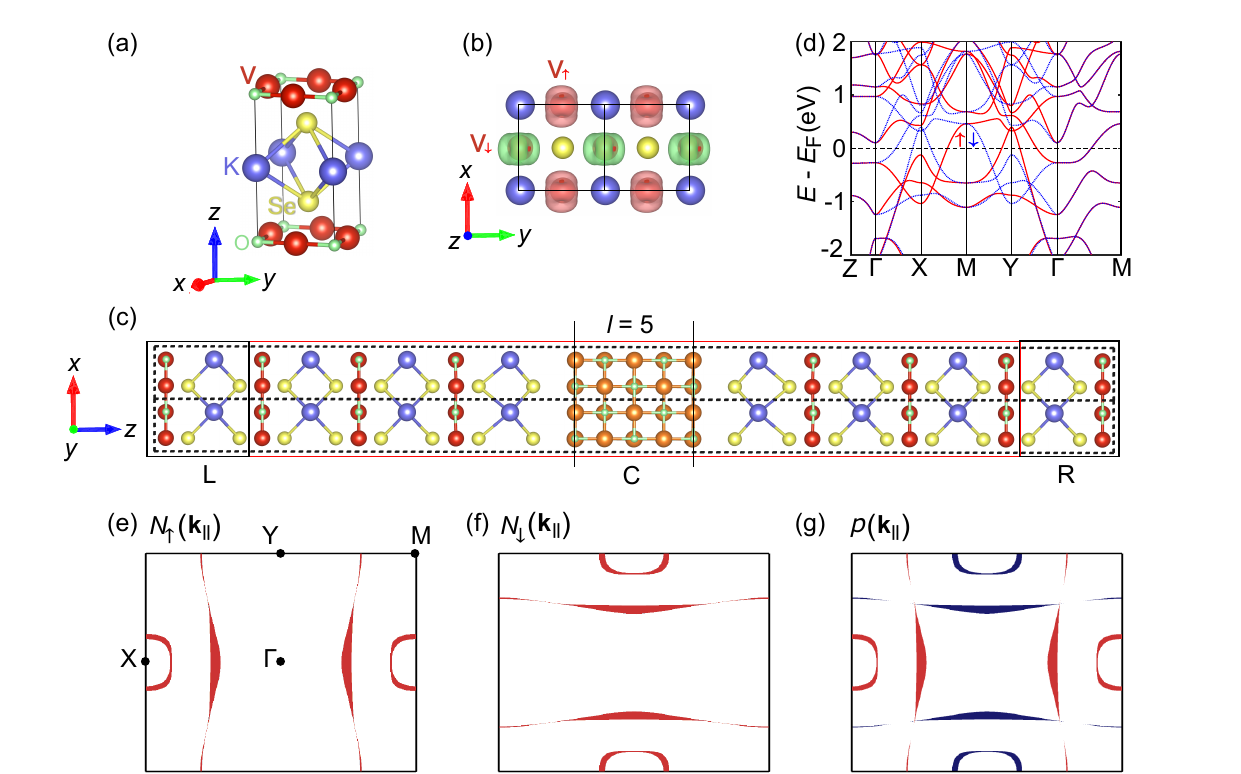}}
 \caption{ \textbf{Geometrical and electronic structures of KV$_2$Se$_2$O/MgO/KV$_2$Se$_2$O MTJs.} (a) Crystalline structure of bulk KV$_2$Se$_2$O. (b) Isosurface plot of the spin-resolved charge densities of bulk KV$_2$Se$_2$O. The isosurface value is set at $3\times10^{-7}$ states/(Hartree$\cdot$Bohr$^3$). (c) Schematic plot of the KV$_2$Se$_2$O/$l$-MgO/KV$_2$Se$_2$O magnetic tunneling junction with $l=5$. Two unit cells along the $x$ direction are shown for better demonstration. (d) Spin-resolved band structure of bulk KV$_2$Se$_2$O, (e-f) Spin-up, spin-down conduction channels $N_{\uparrow}$, $N_{\downarrow}$, and (g) $\textbf{k}_\parallel$-resolved spin polarization of conduction channels $p({\bf k}_\parallel)$ of bulk KV$_2$Se$_2$O illustrated in the first Brillouin Zone. In (e-f), regions rendered in red/white have one/no electronic state. In (g), red and blue colors indicate 100\% and -100\%  spin polarization, respectively.}\label{model}
 \end{figure*}

\section*{Results}
\subsection*{Effective spin polarization}
For a 3D or 2D heterojunction, there are subbands along the transverse direction. For each transverse wavevector ${\bf{k}}_{\parallel}$, we can deal with the transport problem using the folded Hamiltonian\cite{cxb_PRB2019,QiXiaoliang_PRB2008,cxb_NJP2024}
\begin{equation}
    {H_\alpha }\left( {{{\bf{k}}_{\parallel}}} \right) = \sum\nolimits_{{{\bf{R}}_{\parallel}}} {H_\alpha }\left( {{{\bf{R}}_{\parallel}}} \right){e^{i{{\bf{k}}_{\parallel}} \cdot {{\bf{R}}_{\parallel}}}},
\end{equation}
where $\alpha=\textrm{L/C/R}$ indicates the left lead (L), central region, and the right lead (R). ${H_\alpha }( {{{\bf{R}}_{\parallel}}})$ is the interaction between the 0$^{\textrm {th}}$ unit cell and the unit cell with transverse displacement vector ${{\bf{R}}_{\parallel}}$. In this way, the transport system can be treated as a quasi-1D transport problem [Fig.~\ref{theory}(a)]. To guide the search for high-MR junctions, we show that MR can be estimated using the generalized Julliere formula (see Sec.~I.A and Fig.~S1 in the Supporting Information (SI)~\cite{SI}):
\begin{align}\label{eq:MRop}
        {\rm{M}}{{\rm{R}}_{{\rm{o/p}}}} = \frac{{2\left\langle p\right\rangle ^2}}{1 \mp \left\langle p \right\rangle ^2},
\end{align}
where $\left\langle p \right\rangle$ is the effective spin polarization defined as
\begin{align}\label{eq:avep}
\left\langle p \right\rangle  =\sqrt{ \frac{{\sum\nolimits_{{{\bf{k}}_\parallel } \in {U_{{\rm{LR}}}}} {{p_{\rm{L}}({{\bf{k}}_\parallel })}{p_{\rm{R}}({{\bf{k}}_\parallel })}N_{\rm{L}}\left( {{{\bf{k}}_\parallel }} \right)N_{\rm{R}}\left( {{{\bf{k}}_\parallel }} \right)} }}{{\sum\nolimits_{{{\bf{k}}_\parallel } \in {U_{{\rm{LR}}}}} {N_{\rm{L}}\left( {{{\bf{k}}_\parallel }} \right)N_{\rm{R}}\left( {{{\bf{k}}_\parallel }} \right)} }}}.
\end{align}
and $p_\alpha({\bf{k}}_{\|})$ is the $\textbf{k}_\parallel$-resolved spin polarization of transport channels with energy $E_F$ in lead $\alpha$ ($\alpha=$L,R)~\cite{shao2021spin}:
\begin{equation}\label{eq:p}
p_\alpha({\bf{k}}_{\|})=\frac{N_{\alpha\uparrow}({\bf{k}}_{\|})-N_{\alpha\downarrow}(\bf{k}_{\|})}{N_{\alpha\uparrow}({\bf{k}}_{\|})+N_{\alpha\downarrow}({\bf{k}}_{\|})}.
\end{equation}
Here, $N_\alpha( {{{\bf{k}}_\parallel }})$ is the number of transport channels with energy $E_F$ at ${{{\bf{k}}_\parallel }}$ and $N_{\alpha\sigma}( {{{\bf{k}}_\parallel }})$ is the number of spin-$\sigma$ channels. The summation proceeds over ${{{\bf{k}}_\parallel }}$ in union $U_{\textrm {LR}}$, which is the region of transverse $\bf k_{\|}$ vectors with existing transport channels in the first Brillouin zone for both leads L and R. If leads L and R are the same material, Eq.~(\ref{eq:avep}) reduces to
\begin{align}
\left\langle p \right\rangle  = \sqrt {\frac{{\sum\nolimits_{} {{p^2}({{\bf{k}}_\parallel }){N^2}\left( {{{\bf{k}}_\parallel }} \right)} }}{{\sum\nolimits_{} {{N^2}\left( {{{\bf{k}}_\parallel }} \right)} }}},
\end{align}
where $p({{\bf{k}}_\parallel })$ and $N({{\bf{k}}_\parallel })$ are the spin-polarization and number of transport channels at ${{\bf{k}}_\parallel }$ at $E_F$ for the electrode material. If the spin-up and spin-down transport channels do not overlap with $N({{\bf{k}}_\parallel })=1$ and $p({{\bf{k}}_\parallel }=\pm1)$, we have $\left\langle p \right\rangle  = 1$.

Equation~(\ref{eq:MRop}) indicates that materials with $\langle p\rangle=1$ would reach the extreme-limit MR. Such materials exhibit a key characteristic that at each $\bf{k}_\parallel$ spin-up and spin-down channels have no overlap in the first BZ, which can be named as a spin-valley-mismatched (SVM) altermagnets~\cite{yan2025giant}. For $\langle p\rangle > 0.9$,  MR$_{\textrm p}\approx \langle p\rangle$.

We now discuss the physical picture of the limiting-value MR achieved using materials with $\langle p\rangle=1$. As shown in Fig.~\ref{theory}(b), if lead L has only spin-up states and lead R has only spin-down states for one specific ${\bf{k}}_{\parallel}$, transmission of electrons from L to R is prohibited and vice versa (Proof in Sec. I.B in SI). The distribution of spin-up and spin-down states can switch when the N\'eel vector of an antiferromagnet flips. Therefore, an antiferromagnetic material with unoverlapped spin-up and spin-down states can be used as the L and R leads, for which N\'eel vectors are parallel and antiparallel under the P and AP configurations. Electrons in such a junction have high transmission under P configuration but low transmission under AP configuration, leading to extremely large MR.

To achieve high MR,  the electrodes should satisfy: (1) The system is more than 1D. The reason is that spin-up and spin-down transport channels always overlap with each other for 1D systems. (2) Spin-up and spin-down channels have nearly no overlap. Besides these two conditions, layered materials are favored because non-overlapping Fermi surfaces at $k_z=0$ ($k_z$ is along the out-of-plane direction) guarantee non-overlapping for all transport channels. However, it should be noted that different transport directions may lead to completely different results. As shown in Fig.~\ref{theory}(c), for a 3D system with unoverlapped isoenergy surfaces, the transport channels are fully overlapped if the transport is along the $x$ direction and fully separated if the transport is along the $y$ or $z$ directions.

\subsection*{Devices}
For antiferromagnetic materials, spin-valley mismatching suggests the existence of spin splitting, which is a feature owned by altermagnets~\cite{chen2025unconventional,zhang2024predictable,xu2025chemical,karube2022observation,krempasky2024altermagnetic,zhu2024observation,ding2024large,zeng2024observation}. Recently, Jiang \textit{et~al.} identified that KV$_2$Se$_2$O, which hosts a significant spin splitting as high as 1.6~eV at the Fermi surface, is a metallic room-temperature altermagnet with spin-momentum locking~\cite{jiang2025metallic}. As shown in Figure \ref{model}(a), KV$_2$Se$_2$O possesses a tetragonal layered crystalline structure (space group \textit{P}4/\textit{mmm}) with lattice parameters $a=b=3.946$~\AA~and $c=7.312$~\AA~ at 293 K. Its N\'eel vector points along the [001] direction~\cite{jiang2025metallic}. These lattice parameters are adopted to simulate the transport properties at room temperature in this work ~\cite{lin2018structure,bai2024absence}.

Figure \ref{model}(b) displays the spin-resolved charge densities of bulk KV$_2$Se$_2$O. The charge densities of the two spin sublattices of vanadium (V) atoms undergo interconversion through a $90^{\circ}$ rotation around the $c$ axis. Correspondingly, the sublattices are connected by the spin group symmetry $\left[\mathrm{C}_{2}| | \mathrm{C}_{4 z}\right]$, where $\mathrm{C}_2$ is the 180$^\circ$ spin rotation in the spin space and $\mathrm{C}_{4z}$ is the 90$^\circ$ rotational symmetry with respect to the $z$ axis in the real space~\cite{jiang2025metallic}.

For the construction of an MTJ, we choose MgO as the barrier material because its lattice parameters are close to those of KV$_2$Se$_2$O~\cite{sasaki1979x}. Figure \ref{model}(c) illustrates the schematic representation of the transport model, KV$_2$Se$_2$O/5-MgO/KV$_2$Se$_2$O, where 5-layer MgO(001) is positioned between two KV$_2$Se$_2$O(001) terminals. The MTJ is periodic along the $x$ and $y$ directions, and both the L and R leads are crystalline KV$_2$Se$_2$O.

\begin{figure}[ht]
 \centerline{\includegraphics[width=0.45\textwidth]{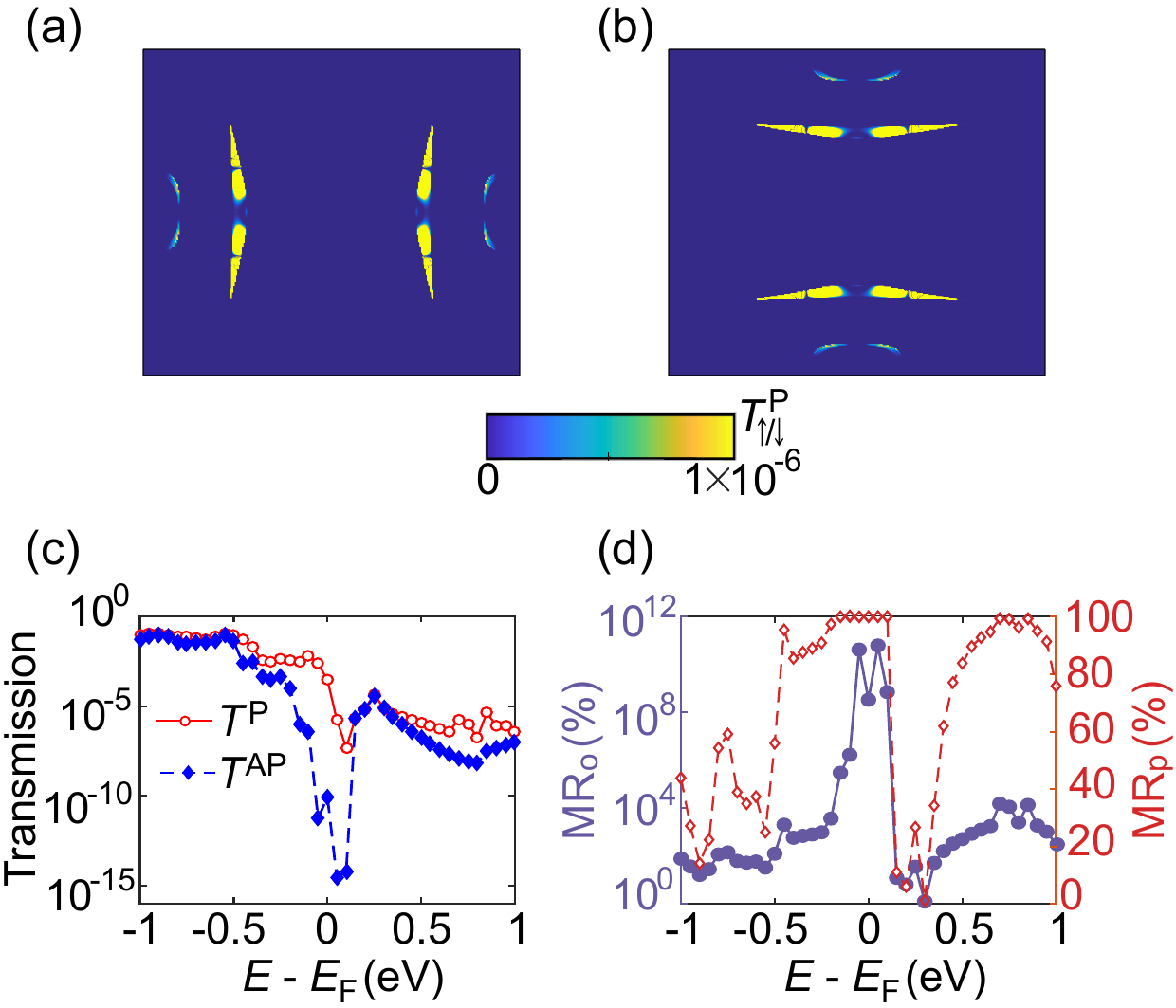}}
 \caption{\textbf{Zero-bias transport properties of the KV$_2$Se$_2$O/5-MgO/KV$_2$Se$_2$O junction.} (a) Spin-up and (b) spin-down transmission spectra under the P configuration. (c) Total transmission coefficients under the P and AP configurations ($T^{\text{P/AP}}$) as a function of energy. (d) MR as a function of energy.
 }\label{transmission}
 \end{figure}

\subsection*{KV$_2$Se$_2$O-based MTJs}
As demonstrated by the band structure in Fig.~\ref{model}(d), KV$_2$Se$_2$O is a metallic antiferromagnet with spin splitting, \textit{i.e.}, an altermagnet. The separation of spin-resolved electronic states at the Fermi energy ($E_F$) is consistent with previous reports~\cite{jiang2025metallic}. Figures~\ref{model}(e-f) further display the number of spin-up (spin-down) transport channels $N_{\uparrow}(N_{\downarrow})$
at $E_F$ for different transverse wave vector $\bf{k}_{\|}$ in the 2D Brillouin zone of crystalline KV$_2$Se$_2$O. It shows that there is either one spin-up or one spin-down channel in the conductive regions. Interestingly, the existing areas of spin-up and spin-down conducting channels in the $\bf k$ space resemble the Fermi surfaces~\cite{jiang2025metallic} because the area of conducting channels is just a projection of the Fermi surface on the transverse plane in the Brillouin zone. Thus, the distribution of $N_{\uparrow}(N_{\downarrow})$  can be connected by a 90$^\circ$ rotational operation, which originates from the features of the band structures. The $\bf{k}_{\|}$-dependent spin polarization is calculated according to Eq.~(\ref{eq:p}) and demonstrated in Fig.~\ref{model}(g). Full spin polarization ($p=\pm 100\%$)
 manifests in nearly the whole conductive region.

From Figs.~\ref{model}(d-g), we know that spin-up and spin-down states of KV$_2$Se$_2$O at $E_F$ barely overlap. To assess the overlapping situation, we calculated the effective spin polarization [Eq.~(\ref{eq:avep})].
It turns out that $\langle p\rangle=99.93$\%, compared to 63\% for RuO$_2$ along the $z$ direction. Although the effective spin polarization of KV$_2$Se$_2$O is nonzero, the spin polarization of KV$_2$Se$_2$O is 0\% due to the degeneracy of spin-up and spin-down bands. Following Eq.~(\ref{eq:MRop}), systems using KV$_2$Se$_2$O as leads can exhibit pessimistic MR as high as $99.93$\%.

\begin{figure}[ht]
 \centerline{\includegraphics[width=0.45\textwidth]{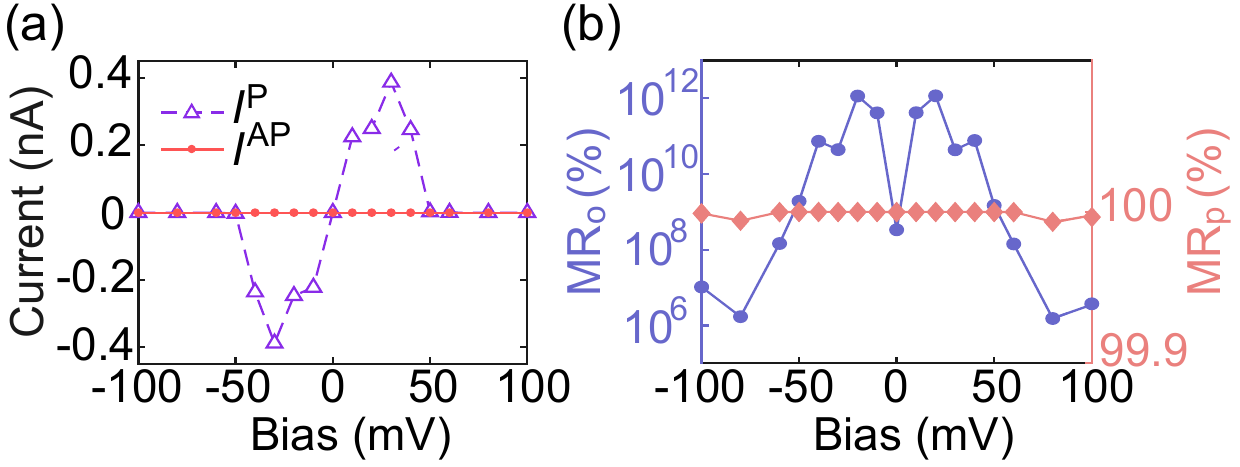}}
 \caption{\textbf{Transport properties of the KV$_2$Se$_2$O/5-MgO/KV$_2$Se$_2$O junction under a bias voltage.} (a) Total current under the P  and AP  configurations. (b) MR as a function of bias voltage. Symbols represent calculated results, and lines are guides to the eye.
 }\label{I-V}
 \end{figure}
For verification, we perform first-principles transport calculations on KV$_2$Se$_2$O/$l$-MgO/KV$_2$Se$_2$O MTJs ($l$-MgO MTJ in short.) using the software NANODCAL~\cite{taylor2001ab,waldron2006first,Yan_Nanotechnology2022,zhanglei_npj2023,yan2025giant}. P and AP alignments are realized by changing the spin polarization of V atoms and thus the N\'{e}el vectors (See Figs.~S2-S2 and Table S1 in SI for structures). We first explore the case of $l=5$. The spin- and ${\bf k}_{\|}$-resolved transmission spectra under parallel configuration of the 5-MgO MTJ at zero bias are depicted in Figs.~\ref{transmission}(a-b). It shows that only parts of the conductive channels can transmit through the heterojunction. As expected, the spin-up and spin-down transmission spectra exhibit a C$_2$ rotational symmetry, which reflects the related spin group symmetry in the distribution of conductive channels in KV$_2$Se$_2$O [Fig.~\ref{model}(b)].

The corresponding zero-bias transmission spectra for the 5-MgO MTJ under both P and AP configurations are demonstrated in Fig.~\ref{transmission}(c) and Fig.~S3. Transmission coefficients of spin-up and spin-down electrons are the same, reflecting the features of antiferromagnetic materials. The transmission coefficients at $E_F$ for the 5-MgO junction are about $2.92\times 10^{-4}$ and $8.66\times 10^{-11}$ for P and AP configurations, respectively, showing a seven-orders-of-magnitude discrepancy in transmission of P and AP configurations. The corresponding zero-bias MR at $E_F$ under P (AP) configurations for the optimistic and pessimistic versions, is as large as $\mathrm{MR}_\textrm{o}=3.37\times 10^{8}\%$ and $\mathrm{MR}_\textrm{p}=99.99997 \%$ (see Sec. III, Fig.~S4, and Tab.~S2 in SI).

The zero-bias MR for the 5-MgO structure is much larger than conventional MTJs using metallic leads, such as 3700\% in Fe/5-MgO/Fe structures~\cite{waldron2006first} and also far exceeds previous theoretical predictions of altermagnet MTJs~\cite{shao2021spin,SmejkalLibor_PRX2022,jiang2023prediction,chi2024crystal,tanaka2024first,chi2025anisotropic,Tsymbal_NL2025}. We also note that the giant tunneling MR is observed not only at the $E_F$ but also at broader energy ranges near $E_F$. As illustrated in Fig.~\ref{transmission}(d), giant tunneling MR is maintained in the vicinity of $E_F$, which indicates that the substantial MR will be maintained within a certain range of bias voltage.

\subsection*{Bias dependence}
The calculated $I$-$V$ curves and MR of the 5-MgO structure as a function of bias are depicted in Fig.~\ref{I-V}. The total current is the sum of spin-up and spin-down currents: $I^{\alpha}=I_{\uparrow}^{\alpha}+I_{\downarrow}^{\alpha}$, \textit{i.e.}, $I^\alpha_\uparrow$ and $I^\alpha_\downarrow$ represent spin-up and spin-down currents at $\alpha$ configuration with $\alpha=\textrm{P, AP}$. In the P configuration, the total current reaches a maximum value at 30 mV. Then, the total current declines and shows a negative differential resistance (NDR) effect. In the AP configuration, the total current is vanishing throughout the range of studied bias voltages [Fig.~\ref{I-V}(a)].
The bias-dependent MR, which can be calculated using the currents under P and AP configuration, is shown in Fig.~\ref{I-V}(b). A maximum MR of $1.2\times 10^{12}\%$ for the 5-MgO junction occurs when a $\pm 20 \mathrm{~mV}$ bias voltage is applied, and the minimum MR of $1.5\times 10^{6}\%$ occurs when a $\pm 80 \mathrm{~mV}$  bias voltage is applied. In other words, MR of the 5-MgO MTJ is indeed robust against small biases (0.1~V) and can be enhanced by nearly 4 orders of magnitude compared to the zero-bias case by applying a small bias voltage of 20 mV.

\begin{figure}[ht]
 \centerline{\includegraphics[width=0.45\textwidth]{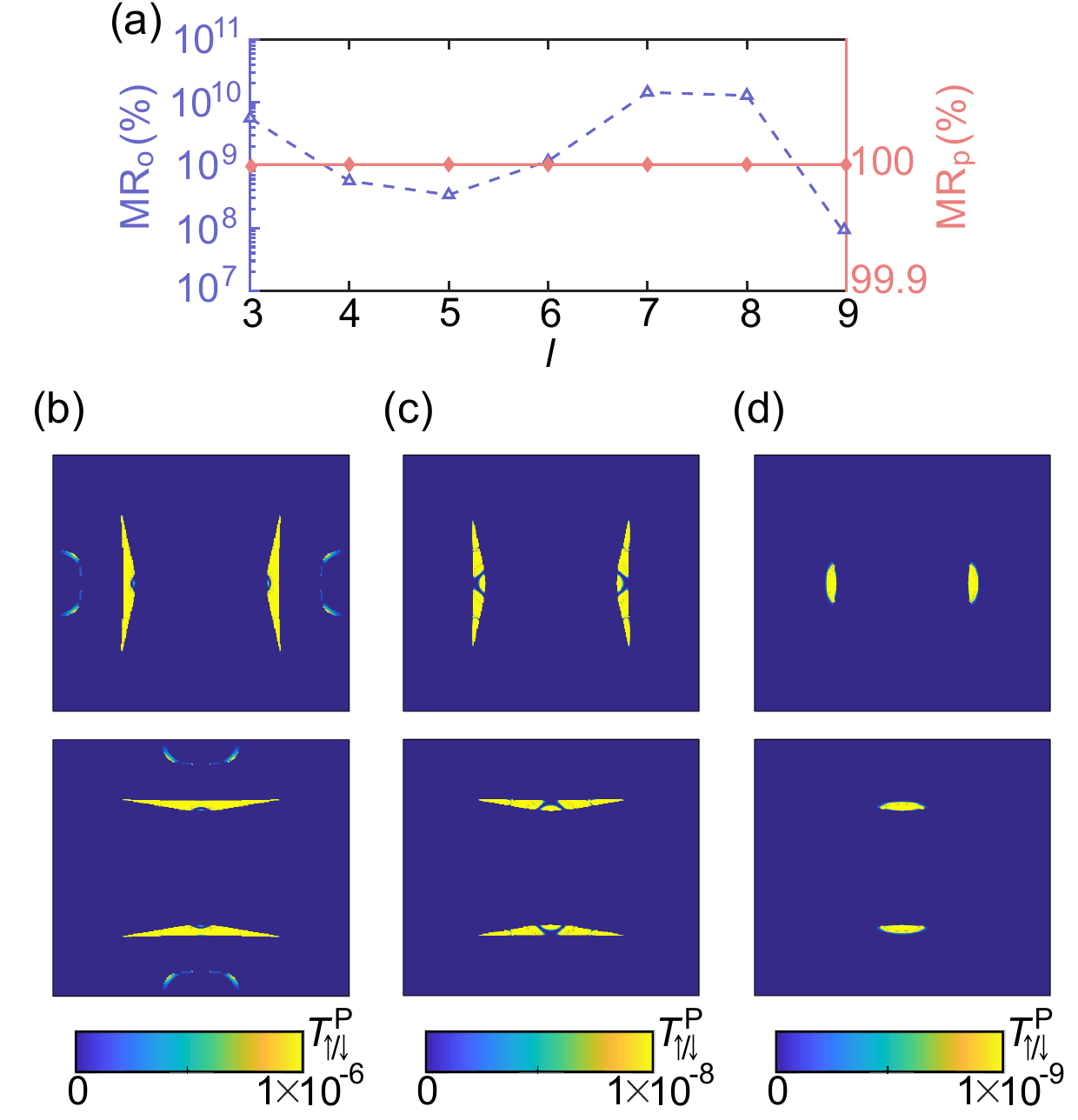}}
 \caption{\textbf{Thickness dependence of KV$_2$Se$_2$O/MgO/KV$_2$Se$_2$O MTJs.} (a) MR of KV$_2$Se$_2$O/$l$-MgO/KV$_2$Se$_2$O junction as a function of the layer number $l$. (b-d) Spin-up (upper panels) and spin-down (lower panels) transmission spectra of KV$_2$Se$_2$O/$l$-MgO/KV$_2$Se$_2$O with $l=3,7,9$ under the P configuration, respectively. }\label{MR-L}
 \end{figure}

\section*{Discussion}
According to Ref.~\onlinecite{yan2025giant}, heterojunctions based on SVM materials, \textit{i.e.}, SVM/Spacer/SVM, exhibit high MR regardless of the spacer materials. However, the thickness of the central barrier region of an MTJ is generally important for the MR. Therefore, we also conducted a systematic study on the variation of zero-bias MR as the number of MgO layers changes from 3 to 9 layers. As shown in Fig.~\ref{MR-L}(a) and Table S2, the maximum MR of $1.44\times 10^{10}\%$ occurs at the heterojunction with 7 layers of MgO, while the minimum MR of $7.57\times 10^{7}\%$ is observed at the heterojunction with 9 layers of MgO. Nevertheless, $l$-MgO MTJ indeed exhibits large MR regardless of the thickness of MgO.

To understand the variation of MR, we also plot the zero-bias spin- and ${\bf k}_{\|}$-resolved transmission spectra of the
$l$-MgO MTJ with $l = 3, 5$, and $9$ in Figs.~\ref{MR-L}(b-d).
We can observe that although the general features of transmission spectra do not change, the distribution of the transmission spectra varies. Specifically, the distribution of the spin-up and spin-down transmission spectra under P configuration is primarily along the perpendicular bisectors of the $\Gamma$-X and $\Gamma$-Y lines. While, the transmission near the X and Y points, which are contributed by $d_{xy}$, $d_{x^2-y^2}$, and $d_{z^2}$ orbitals, is vanishing [compared to Fig.~\ref{MR-L}(b). See Sec. IV and Fig.~S5 in SI]. Furthermore, as the number of MgO barrier layers $l$ increases, both the distribution and magnitudes of the transmission spectra shrink significantly, and the transmission spectra increasingly concentrate around the midpoints of the $\Gamma$-X(Y) line. The increasing concentration can be understood using the simple tunneling model where electrons tunnel through a barrier with height $V_b$ and the transmission spectra scale as $\exp(-2d\kappa)$ with $\kappa^2=(2m/\hbar^2)(V_b-E_F)+k_{\|}^2$ and ${k}_{\|}$ the norm of the transverse $\bf{k}_\|$ vector~\cite{Bulter_PRB2001}.

From the theoretical perspective, the high MR originates from the intrinsic band-structure features of the electrodes; therefore, barrier materials can be arbitrary nonmagnetic materials~\cite{yan2025giant}. To justify this inference, we adopted three different barrier materials, including vacuum with a thickness of 7 \AA,  CaTiO$_3$, and BaTiO$_3$, to construct three KV$_2$Se$_2$O-based MTJs and performed transport calculations. All junctions exhibit high MR larger than $10^6$\% (optimistic MR) and 99.997\% (pessimistic MR) as expected. Details can be found in Sec. V, Figs.~S6-S8, and Table S3 in SI.

Our research focuses on perfect structures. In reality, interfacial defects such as oxygen vacancies can occur and significantly reduce MR~\cite{KeYouqi_PRL2007}. Van der Waals materials would be a better choice for  barrier materials. In addition, both lattice parameters and atomic positions change under different temperatures, which may lower the MR of KV$_2$Se$_2$O-based devices. According to a recent study based on the density functional theory combined with dynamical mean-field theory~\cite{TianFuyang2025}, the room-temperature Fermi surface of KV$_2$Se$_2$O change a little compared with that of low-temperature ones. The effective spin polarization of KV$_2$Se$_2$O and thus the MR of KV$_2$Se$_2$O-based devices are expected to maintain a large value at room temperature.

In summary, we propose an intuitive and predictive theory for the estimation of MR in MTJs, ${\textrm {MR}}_{\textrm {o/p}}= 2\left\langle p \right\rangle^2 /(1 \mp \left\langle p \right\rangle^2)$, which uses the effective spin polarization in the $k$ space. From this equation, we predict that materials with full spin-valley mismatch would exhibit extreme-limit MR. Combining with first-principles calculations, we report that the newly fabricated altermagnet metal KV$_2$Se$_2$O is a spin-valley-mismatched altermagnet. Further, using $l$-layer MgO as the spacer, we built magnetic tunnel junctions KV$_2$Se$_2$O/$l$-MgO/KV$_2$Se$_2$O ($3\le l \le 9$) and explored the magnetoresistive properties by using first-principles transport calculations. It shows that such heterojunctions exhibit zero-bias magnetoresistance larger than $7.57\times10^7\sim10^{10}$\% (optimistic definition) and 99.9998\% (pessimistic definition), which are robust against small biases and variation of spacer thickness. Our research provides a quantitative design principle for next-generation spin-electronic devices and establishes KV$_2$Se$_2$O/MgO/KV$_2$Se$_2$O as a leading candidate material system for room-temperature ultra-high-density non-volatile memory.

\section*{Methods}
\subsection*{Optimization}
In our model, MgO is stretched by 6\% to match the lattice of KV$_2$Se$_2$O. The central scattering area comprises six KV$_2$Se$_2$O layers functioning as buffer layers at both ends, along with $l$-layer MgO. Structures were optimized using the Vienna ab initio simulation package (VASP)~\cite{kresse1996efficient} based on density functional theory (DFT).
The exchange-correlation energy is described by the Perdew-Burke-Ernzerhof (PBE) functional under the generalized gradient approximation (GGA) ~\cite{blochl1994projector}. A vacuum layer larger than 10~\AA~is added in the stacking direction, and the optB86-vdW functional~\cite{klimevs2011van} incorporates the van der Waals (vdW) interaction in the structure. During the structural optimization, the energy convergence is set to $10^{-5}$ eV, and the convergence criterion for forces on each atom is less than 0.03 eV/\AA. The plane-wave cutoff energy is set to 500 eV, and the sampling of the Brillouin zone is based on the Monkhorst-Pack method with a $k$-point grid of $8\times8\times1$ to ensure convergence.

\subsection*{Transport}

Transport properties were calculated using the nonequilibrium Green's function in combination with the density functional theory (NEGF-DFT), as implemented within the NANODCAL package~\cite{taylor2001ab,waldron2006first}. A $k$-point mesh of $16\times 16\times 1$ was employed in the self-consistent calculations of the central region interfaced with two electrodes. In addition, a $160\times 160\times 1$ $k$-mesh grid was used to calculate the transmission coefficients of these models. The bias voltage $V$
is established by setting the chemical potential on the left (right) electrode to $+V/ 2\left(-V / 2\right)$.

\section*{Data availability}
The data supporting this study are available in the GitHub repository at \url{https://github.com/kYan-CMT/Model-of-KV2Se2-MgO-kVSe2O.git}.


\section*{Acknowledgements}
We gratefully acknowledge the financial support from the Natural Science Foundation of China (Grants Nos. 12574256 and 12447144) and the Shenzhen Science and Technology Program (Grant Nos. RCYX20221008092848063, ZDSYS20230626091100001, JCYJ20241202123733043, and JCYJ20241202123506009) and the Shenzhen Basic Research Projects (No. JCYJ20241202123916023).\\


\section*{Author contributions}
X.C. developed the theory and guided this work. K.Y. performed first-principles transport calculations. K.Y., X.C., and X.Z. wrote the manuscript. Y.H., W.X., and W.D. discussed and revised the manuscript.

\section*{Competing interests}
The authors declare no competing financial or non-financial interests.

A pending Chinese patent application (Application No. 2026100488624) is relevant to this work. Applicant: Harbin Institute of Technology, Shenzhen; Inventors: Kun Yan, Xiaobin Chen; Status: Pending (preliminary examination passed). The patent covers the structure and materials described in Figs.4(b) and 5(a).

Author Wenhui Duan is an editorial board member of npj Computational Materials. Wenhui Duan was not involved in the journal’s review of, or decisions related to, this manuscript.

The other authors do not have a competing interest.

\section*{Additional information}
\textbf{Supplementary information} The online version contains supplementary material available at https://***

\textbf{Correspondence} and requests for materials should be addressed to Xiaolong Zou and Xiaobin Chen.

\section*{References}

\begin{thebibliography}{10}

\bibitem{julliere1975tunneling}
Julliere, M.
\newblock {\em Phy. Lett. A}{ \bf 54}(3), 225--226 (1975).

\bibitem{yuasa2004giant}
Yuasa, S., Nagahama, T., Fukushima, A., Suzuki, Y., and Ando, K.
\newblock {\em Nat. Mater.}{ \bf 3}(12), 868--871 (2004).

\bibitem{moodera1995large}
Moodera, J.~S., Kinder, L.~R., Wong, T.~M., and Meservey, R.
\newblock {\em Phys. Rev. Lett.}{ \bf 74}(16), 3273 (1995).

\bibitem{wang2015spin}
Wang, W., Narayan, A., Tang, L., Dolui, K., Liu, Y., Yuan, X., Jin, Y., Wu, Y., Rungger, I., Sanvito, S., et~al.
\newblock {\em Nano Lett.}{ \bf 15}(8), 5261--5267 (2015).

\bibitem{zhao2017magnetic}
Zhao, K., Xing, Y., Han, J., Feng, J., Shi, W., Zhang, B., and Zeng, Z.
\newblock {\em J. Magn. Magn. Matter.}{ \bf 432}, 10--13 (2017).

\bibitem{jiang2023prediction}
Jiang, Y.-Y., Wang, Z.-A., Samanta, K., Zhang, S.-H., Xiao, R.-C., Lu, W., Sun, Y., Tsymbal, E.~Y., and Shao, D.-F.
\newblock {\em Phys. Rev. B}{ \bf 108}(17), 174439 (2023).

\bibitem{naka2019spin}
Naka, M., Hayami, S., Kusunose, H., Yanagi, Y., Motome, Y., and Seo, H.
\newblock {\em Nat. Commun.}{ \bf 10}(1), 4305 (2019).

\bibitem{vzelezny2017spin}
{\v{Z}}elezn{\`y}, J., Zhang, Y., Felser, C., and Yan, B.
\newblock {\em Phys. Rev. Lett.}{ \bf 119}(18), 187204 (2017).

\bibitem{gurung2021transport}
Gurung, G., Shao, D.-F., and Tsymbal, E.~Y.
\newblock {\em Phys. Rev. Mater}{ \bf 5}(12), 124411 (2021).

\bibitem{vsmejkal2022beyond}
{\v{S}}mejkal, L., Sinova, J., and Jungwirth, T.
\newblock {\em Phys. Rev. X}{ \bf 12}(3), 031042 (2022).

\bibitem{vsmejkal2022emerging}
{\v{S}}mejkal, L., Sinova, J., and Jungwirth, T.
\newblock {\em Phys. Rev. X}{ \bf 12}(4), 040501 (2022).

\bibitem{bai2022observation}
Bai, H., Han, L., Feng, X., Zhou, Y., Su, R., Wang, Q., Liao, L., Zhu, W., Chen, X., Pan, F., et~al.
\newblock {\em Phys. Rev. Lett.}{ \bf 128}(19), 197202 (2022).

\bibitem{shao2021spin}
Shao, D.-F., Zhang, S.-H., Li, M., Eom, C.-B., and Tsymbal, E.~Y.
\newblock {\em Nat. Commun.}{ \bf 12}(1), 7061 (2021).

\bibitem{chi2024crystal}
Chi, B., Jiang, L., Zhu, Y., Yu, G., Wan, C., Zhang, J., and Han, X.
\newblock {\em Phys. Rev. Appl}{ \bf 21}(3), 034038 (2024).

\bibitem{xu2023spin}
Xu, S., Zhang, Z., Mahfouzi, F., Huang, Y., Cheng, H., Dai, B., Kim, J., Zhu, D., Cai, W., Shi, K., Guo, Z., Cao, K., Hong, B., Liu, Y., Yang, J., Zhang, K., Cao, J., Zhu, F., Tai, L., Wang, Y., Eimer, S., Fert, A., Wang, K.~L., Kioussis, N., Zhang, Y., and Zhao, W.
\newblock {\em Nat. Commun.}{ \bf 16}(1), 8370 (2025).

\bibitem{Noh2025_RuO2TMR}
Noh, S., Kim, G.-H., Lee, J., Jung, H., Seo, U., So, G., Lee, J., Lee, S., Park, M., Yang, S., Oh, Y.~S., Jin, H., Sohn, C., and Yoo, J.-W.
\newblock {\em Phys. Rev. Lett.}{ \bf 134}(24), 246703 (2025).

\bibitem{Wadley_Science2016}
Wadley, P., Howells, B., \v{Z}elezny, J., Andrews, C., Hills, V., Campion, R.~P., Novak, V., Olejnik, K., Maccherozzi, F., Dhesi, S.~S., Martin, S.~Y., Wagner, T., Wunderlich, J., Freimuth, F., Mokrousov, Y., Kunes, J., Chauhan, J.~S., Grzybowski, M.~J., Rushforth, A.~W., Edmonds, K.~W., Gallagher, B.~L., and Jungwirth, T.
\newblock {\em Science}{ \bf 351}(6273), 587--590 (2016).

\bibitem{LuHaizhou_PRL2025}
Chen, Y., Liu, X., Lu, H.-Z., and Xie, X.~C.
\newblock {\em Phys. Rev. Lett.}{ \bf 135}(1), 016701 (2025).

\bibitem{Taoll_PRL2024}
Tao, L.~L., Zhang, Q., Li, H., Zhao, H.~J., Wang, X., Song, B., Tsymbal, E.~Y., and Bellaiche, L.
\newblock {\em Phys. Rev. Lett.}{ \bf 133}(9), 096803 (2024).

\bibitem{fu2025}
Fu, P.-H., Lv, Q., Xu, Y., Cayao, J., Liu, J.-F., and Yu, X.-L.
\newblock {\em npj Quantum Mater.}{ \bf 10}(1), 111 (2025).

\bibitem{SmejkalLibor_PRX2022}
\v{S}mejkal, L., Hellenes, A.~B., Gonzalez-Hernandez, R., Sinova, J., and Jungwirth, T.
\newblock {\em Phys. Rev. X}{ \bf 12}(1), 011028 (2022).

\bibitem{cxb_PRB2019}
Chen, X., Xu, Y., Wang, J., and Guo, H.
\newblock {\em Phys. Rev. B}{ \bf 99}(6), 064302 (2019).

\bibitem{QiXiaoliang_PRB2008}
Qi, X.-L., Hughes, T.~L., and Zhang, S.-C.
\newblock {\em Phys. Rev. B}{ \bf 78}(19), 195424 (2008).

\bibitem{cxb_NJP2024}
Chen, X., Cui, S., Hu, Y., and Yan, K.
\newblock {\em New J. Phys.}{ \bf 26}(10), 103035 (2024).

\bibitem{SI}
See the Supplemental Information at http://*** for (1) the theory of estimating MR using the generalized spin polarization and prohibited transport under the AP configuration for spin-valley-mismatched altermagnet, (2) structures and magnetic configuration setup, (3) zero-bias transmission spectra of KV$_2$Se$_2$O/$l$-MgO/KV$_2$Se$_2$O, (4) projected density of states of V atoms in bulk KV$_2$Se$_2$O, and (5) influence of barrier materials in KV$_2$Se$_2$O-based MTJs, which includes Refs.~\cite{cxb_PRB2019,ALI20052867,al2011three}.

\bibitem{yan2025giant}
Yan, K., Cheng, L., Hu, Y., Gao, J., Zou, X., and Chen, X.
\newblock {\em Phys. Rev. Lett.}{ \bf 134}(3), 036302 (2025).

\bibitem{chen2025unconventional}
Chen, X., Liu, Y., Liu, P., Yu, Y., Ren, J., Li, J., Zhang, A., and Liu, Q.
\newblock {\em Nature}{ \bf 640}, 349--354 (2025).

\bibitem{zhang2024predictable}
Zhang, R.-W., Cui, C., Li, R., Duan, J., Li, L., Yu, Z.-M., and Yao, Y.
\newblock {\em Phys. Rev. Lett.}{ \bf 133}(5), 056401 (2024).

\bibitem{xu2025chemical}
Xu, R., Gao, Y., and Liu, J.
\newblock {\em Natl. Sci. Rev.}{ \bf 13}(2), nwaf528 (2026).

\bibitem{karube2022observation}
Karube, S., Tanaka, T., Sugawara, D., Kadoguchi, N., Kohda, M., and Nitta, J.
\newblock {\em Phys. Rev. Lett.}{ \bf 129}(13), 137201 (2022).

\bibitem{krempasky2024altermagnetic}
Krempask{\`y}, J., {\v{S}}mejkal, L., D'souza, S., Hajlaoui, M., Springholz, G., Uhl{\'\i}{\v{r}}ov{\'a}, K., Alarab, F., Constantinou, P., Strocov, V., Usanov, D., et~al.
\newblock {\em Nature}{ \bf 626}(7999), 517--522 (2024).

\bibitem{zhu2024observation}
Zhu, Y.-P., Chen, X., Liu, X.-R., Liu, Y., Liu, P., Zha, H., Qu, G., Hong, C., Li, J., Jiang, Z., et~al.
\newblock {\em Nature}{ \bf 626}(7999), 523--528 (2024).

\bibitem{ding2024large}
Ding, J., Jiang, Z., Chen, X., Tao, Z., Liu, Z., Li, T., Liu, J., Sun, J., Cheng, J., Liu, J., et~al.
\newblock {\em Phys. Rev. Lett.}{ \bf 133}(20), 206401 (2024).

\bibitem{zeng2024observation}
Zeng, M., Zhu, M.-Y., Zhu, Y.-P., Liu, X.-R., Ma, X.-M., Hao, Y.-J., Liu, P., Qu, G., Yang, Y., Jiang, Z., et~al.
\newblock {\em Adv. Sci.}{ \bf 11}(43), 2406529 (2024).

\bibitem{jiang2025metallic}
Jiang, B., Hu, M., Bai, J., Song, Z., Mu, C., Qu, G., Li, W., Zhu, W., Pi, H., Wei, Z., et~al.
\newblock {\em Nat. Phys.}{ \bf 21}, 754 (2025).

\bibitem{lin2018structure}
Lin, H., Si, J., Zhu, X., Cai, K., Li, H., Kong, L., Yu, X., and Wen, H.-H.
\newblock {\em Phys. Rev. B}{ \bf 98}(7), 075132 (2018).

\bibitem{bai2024absence}
Bai, J., Ruan, B., Dong, Q., Zhang, L., Liu, Q., Cheng, J., Liu, P., Li, C., Sun, Y., Huang, Y., et~al.
\newblock {\em Phys. Rev. B}{ \bf 110}(16), 165151 (2024).

\bibitem{sasaki1979x}
Sasaki, S., Fujino, K., and Tak{\'e}uchi, Y.
\newblock {\em Pro. Jpn. Acd. B}{ \bf 55}(2), 43--48 (1979).

\bibitem{taylor2001ab}
Taylor, J., Guo, H., and Wang, J.
\newblock {\em Phys. Rev. B}{ \bf 63}(24), 245407 (2001).

\bibitem{waldron2006first}
Waldron, D., Timoshevskii, V., Hu, Y., Xia, K., and Guo, H.
\newblock {\em Phys. Rev. Lett.}{ \bf 97}(22), 226802 (2006).

\bibitem{Yan_Nanotechnology2022}
Yan, K., Hu, Y., Suo, Y., Qin, Y., and Chen, X.
\newblock {\em Nanotechnology}{ \bf 33}(38), 385001 (2022).

\bibitem{zhanglei_npj2023}
Xiao, W., Zheng, X., Hao, H., Kang, L., Zhang, L., and Zeng, Z.
\newblock {\em npj Comput. Mater.}{ \bf 9}(1), 144 (2023).

\bibitem{tanaka2024first}
Tanaka, K., Nomoto, T., and Arita, R.
\newblock {\em Phys. Rev. B}{ \bf 110}(6), 064433 (2024).

\bibitem{chi2025anisotropic}
Chi, B., Jiang, L., Zhu, Y., Yu, G., Wan, C., and Han, X.
\newblock {\em Phys. Rev. Appl.}{ \bf 23}(1), 014013 (2025).

\bibitem{Tsymbal_NL2025}
Samanta, K., Shao, D.-F., and Tsymbal, E.~Y.
\newblock {\em Nano Lett.}{ \bf 25}(8), 3150--3156 (2025).

\bibitem{Bulter_PRB2001}
Butler, W.~H., Zhang, X.~G., Schulthess, T.~C., and MacLaren, J.~M.
\newblock {\em Phys. Rev. B}{ \bf 63}(5), 054416 (2001).

\bibitem{KeYouqi_PRL2007}
Ke, Y., Xia, K., and Guo, H.
\newblock {\em Phys. Rev. Lett.}{ \bf 105}(23), 236801 (2010).

\bibitem{TianFuyang2025}
Xu, Y., Zhang, H., Feng, M., and Tian, F.
\newblock {\em Phys. Rev. B}{ \bf 112}(12), 125141 (2025).

\bibitem{kresse1996efficient}
Kresse, G. and Furthm{\"u}ller, J.
\newblock {\em Phys. Rev. B}{ \bf 54}(16), 11169 (1996).

\bibitem{blochl1994projector}
Bl{\"o}chl, P.~E.
\newblock {\em Phys. Rev. B}{ \bf 50}(24), 17953 (1994).

\bibitem{klimevs2011van}
Klime{\v{s}}, J., Bowler, D.~R., and Michaelides, A.
\newblock {\em Phys. Rev. B}{ \bf 83}(19), 195131 (2011).

\bibitem{ALI20052867}
Ali, R. and Yashima, M.
\newblock {\em Solid State Chem.}{ \bf 178}(9), 2867--2872 (2005).

\bibitem{al2011three}
Al-Shakarchi, E.~K. and Mahmood, N.~B.
\newblock {\em Mod. Phys.}{ \bf 2}(11), 1420--1428 (2011).

\end{thebibliography}

\end{document}